# Emulation methods and adaptive sampling increase the efficiency of sensitivity analysis for computationally expensive models

**Authors:** Haochen Ye[1], Robert E. Nicholas[2,3], Vivek Srikrishnan[4], and Klaus Keller[5]

[1]Department of Geosciences, The Pennsylvania State University, University Park, PA, USA

[2]Earth and Environmental System Institute, The Pennsylvania State University, University Park, PA, USA

[3]Department of Meteorology and Atmospheric Science, The Pennsylvania State University, University Park, PA, USA

[4]Department of Biological and Environmental Engineering, Cornell University, Ithaca, NY, USA

[5]Thayer School of Engineering at Dartmouth College, Hanover, NH, USA

## Abstract

Models with high-dimensional parameter spaces are common in many applications. Global sensitivity analyses can provide insights on how uncertain inputs and interactions influence the outputs. Many sensitivity analysis methods face nontrivial challenges for computationally demanding models. Common approaches to tackle these challenges are to (i) use a computationally efficient emulator and (ii) sample adaptively. However, these approaches still involve potentially large computational costs and approximation errors. Here we compare the results and computational costs of four existing global sensitivity analysis methods applied to a test problem. We sample different model evaluation time and numbers of model parameters. We find that the emulation and adaptive sampling approaches are faster than Sobol' method for slow models. The Bayesian adaptive spline surface method is the fastest for most slow and high-dimensional models. Our results can guide the choice of a sensitivity analysis method under computational resources constraints.

# 1. Introduction

Complex models involving multiple inputs with large uncertainties are common in many applications such as climate risk management and Earth system modeling (Anderies et al. 2007; Roe and Baker 2007; Adger et al. 2018; Keller et al. 2021). Large parametric uncertainties often lead to considerable uncertainty surrounding projections which can complicate decision-making (de Moel et al. 2012; Wong et al. 2018; Zarekarizi et al. 2020; Beevers et al. 2020). Understanding the relative importance of parameters in driving output uncertainty can help to guide refined uncertainty characterizations and assessments of which uncertainties drive poor outcomes (Saltelli 2002; Pianosi and Wagener 2016; Zarekarizi et al. 2020).

Sensitivity analysis is a common tool for quantifying how much of the output uncertainty can be explained by the data, forcing, input parameters and their interactions (Reed et al., 2022). However, the results of sensitivity analysis can hinge on the choice of sensitivity analysis method, the sampling strategy, and the number of model evaluations (Iooss and Saltelli 2016; Pianosi et al. 2016; Sarrazin et al. 2016; Wagener and Pianosi 2019). Approaches to sensitivity analyses include local and global methods, where local sensitivity analysis varies model parameters within a particular reference range, and global sensitivity analysis allows varying all the input parameters within a reasonable input space (Homma and Saltelli 1996; Iooss and Saltelli 2016). Global sensitivity analysis is crucial to the analysis of the potential nonlinear effects among multiple uncertain parameters in the entire input space (Wagener and Pianosi 2019).

The Sobol' method is a common approach to variance-based global sensitivity analysis (Sobol' 2001). The Sobol' method provides a straightforward way to quantify and prioritize input parameters and interactions. However, when dealing with models with a high dimensional space of uncertain parameters, the Sobol' method typically requires a relatively large number of samples (often exceeding tens of thousands) for the sensitivity indices to converge. This computational requirement may be infeasible for high-dimensional or computationally expensive models (Saltelli 2002; Sarrazin et al. 2016).

Emulation provides one approach to mitigate this problem (Ratto et al. 2012). The emulation approach first uses a small number of model runs to fit a surrogate model to approximate the response surface of the original expensive model. This emulator, which is designed to be less computationally expensive than the original model, is then used for the sensitivity analyses. This approach can reduce the computational burden imposed by the expensive models, though the accuracy of the sensitivity analysis results depends on the quality of the emulator.

Adaptive sampling is another approach to reduce the computational burden (Jones et al. 1998; Busby 2009; Garud et al. 2017). This approach involves an active learning step that evaluates a learning function to choose the next sample, which can increase the efficiency of the emulation process. For example, adaptive sampling methods have been shown to be efficient for structural reliability analysis and computational experimental design (Echard et al. 2011; Lookman et al. 2019; Fuhg et al. 2021).

However, it is not guaranteed that emulation and/or adaptive sampling outperform the standard Sobol' approach in terms of computational costs. This is because the emulation process and learning process both take extra time, and the emulation step can introduce approximation errors. The computational costs of these processes often grow nonlinearly as the model dimension increases (Shan and Wang 2010). Moreover, when emulating high-dimensional models, a larger sample size is often required to control the approximation errors within an acceptable range (Liu and Guillas 2017).

Choosing a sensitivity analysis method and sampling strategy to produce sufficiently accurate sensitivity indices with manageable computational costs can pose a difficult decision problem. Understanding the relationship between each method's computational costs and model characteristics such as model parameter dimension and evaluation time can help with this decision. Previous studies have broken important new ground in this area by summarizing and comparing the convergence rates and computational costs of some sensitivity analysis methods in a range of applications (Frey and Patil 2002; Herman et al. 2013; Vanrolleghem et al. 2015; Sarrazin et al. 2016). Here we expand on this state-of-the art by comparing four commonly-used methods for global sensitivity analysis: (i) the Sobol' method (Sobol' 2001), (ii) the ordinary Kriging emulator method (Oliver and Webster 1990; Kaymaz 2005), (iii) Bayesian Adaptive Spline Surface (BASS) (Francom et al. 2018; Francom and Sansó 2020), and (iv) the adaptive Kriging method combined with Monte Carlo Sampling (AKMCS) (Echard et al. 2011). We adopt the Sobol' method as a base-case and standard given its popularity. The Kriging method is a representative and simple emulation approach widely applied in many areas. The BASS method is a special emulation approach with unique advantages in calculating the sensitivity indices. The AKMCS method is one of the simplest adaptive sampling approaches that expands on the Kriging method. We choose a test function with a polynomial structure that can be easily expanded into multiple dimension spaces. We address two questions: (i) What is the fastest sensitivity analysis method for a given model? (ii) How does the estimated computational cost change?

This paper is organized as follows: in section 2 we introduce the basic settings of the perfect model experiments; in section 3 we introduce the sensitivity analysis methods in more detail; in section 4 we report the main results; in section 5 we discuss some potential limitations; in section 6 we conclude the paper and discuss possible future research directions.

## 2. Experimental setup

Our model design covers different model dimensions (i.e., the number of parameters) and different model evaluation times. We only consider deterministic models with single scalar output for simplicity. We adopt a simple polynomial model structure that can be easily expanded into different dimensions:

$$y = \sum_{i=1}^{n}(x_i + 10x_{i+1}^2 x_i + 100x_{i+2}) \quad (x \in [0,1]). \tag{1}$$

This model repeats its structure pattern every three terms (Equation 1), where every first term ($1^{st}$, $4^{th}$, $7^{th}$, etc.) being a weak linear effect, every second term ($2^{nd}$, $5^{th}$, $8^{th}$, etc.) being a medium interaction effect, and every third term ($3^{rd}$, $6^{th}$, $9^{th}$, etc.) being a strong linear effect. We define all parameters to have the same input range between zero and one for sampling convenience. We expect to see similarly large sensitivity indices for every third parameter because these terms have the largest coefficients. This design simplifies the interpretation of the results. We discuss the choice of the test model in more detail in section 3.

We consider models of six different dimensions: 2, 5, 10, 15, 20, and 30. This covers the range covered by many simple to medium-complexity models (e.g. Hall J. W. et al. 2005; Baroni and Tarantola 2014; Wong and Keller 2017)). For higher-complexity models with many parameters (say, more than 100), it is often infeasible to consider all the parameters at the same time due to the computational constraints. Dealing with these complex models often requires a pre-selection or a dimension-reduction process (Liu and Guillas 2017). For each model dimension, we keep the first corresponding number of terms of the model structure. This gives the corresponding number of parameters in the model. Equation 2 provides an example for a 5-dimensional model:

$$y = x_1 + 10x_1x_2^2 + 100x_3 + x_4 + 10x_4x_5^2. \tag{2}$$

We consider a series of model evaluation times from one microsecond to one day (1$\mu$s, 10$\mu$s, 0.1ms, 1ms, 10ms, 0.1s, 1s, 10s, 1min, 1h, 6h, 12h, 1 day). For each test model, we record its mean evaluation time under the same computational condition. We pad the total computational time by the time difference between the true evaluation time and the considered evaluation time.

For each combination scenario of model dimension and model evaluation time, we perform sensitivity analysis using the four considered methods and record the total required computational time for each method. The total time is the sum of model evaluation time, emulator fitting time (if needed), and sensitivity analysis time.

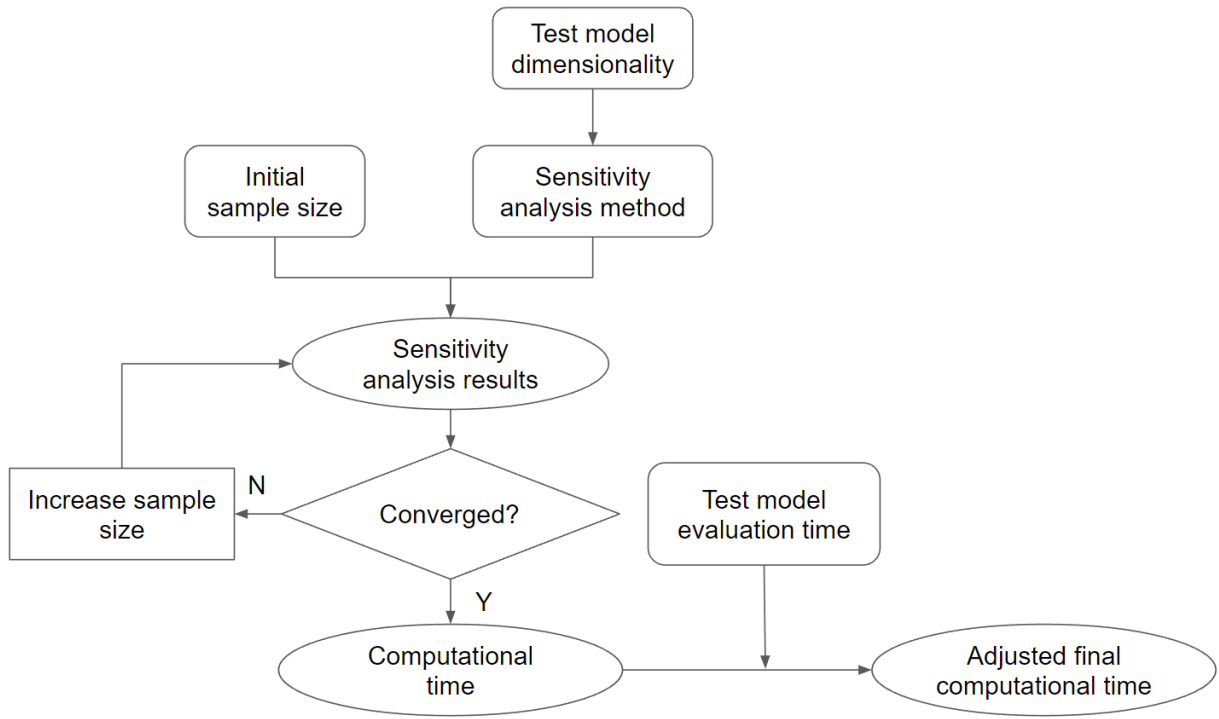

**Figure 1: Analysis flow diagram**

## 3. Sensitivity analysis methods

We consider four sensitivity analysis methods using or expanding on the Sobol' method, which are described further in this section. These methods are: (i) the Sobol' method (Sobol' 2001), (ii) ordinary Kriging (Oliver and Webster 1990; Kaymaz 2005), (iii) Bayesian Adaptive Spline Surface (BASS) (Francom et al. 2018; Francom and Sansó 2020), and (iv) adaptive Kriging method combined with Monte Carlo Sampling (AKMCS) (Echard et al. 2011).

The first method is the standard Sobol' method (Sobol' 2001). The remaining three are different emulation-based methods, where we first build an emulator and then apply Sobol' method to it. Building an emulator often requires far fewer samples over the original model compared with the standard Sobol' analysis. For computationally expensive models, it is more efficient to perform the sensitivity analysis on the computationally emulators, which means emulation-based methods can save time in the model evaluation process. We choose the Sobol' method as our standard and evaluate the performances of the other three methods relative to this standard because of its popularity and interpretability of sensitivity indices as a measure of factor priority. We record each method's final sensitivity analysis results, the number of model evaluations and the total computational time. We do not consider parallelization in this study due to simplicity. The effect of parallelization on model evaluation can be approximately treated as a linear reduction in the total computational time depending on the number of cores in use. We describe the four methods in the next sections.

## 3.1. Sobol' method

Sobol' method is a typical variance-based method, which uses variances to describe the output uncertainty and quantify the importance of each input and each interaction. When all the inputs are independent of each other, Sobol' method decomposes the output variance by a sum of variances explained by each parameter and each interaction:

$$V(y) = \sum_{i=1}^{k} V_i + \sum_{i=1}^{k-1} \sum_{i<j} V_{ij} + \ldots + V_{1,2,\ldots,k}, \qquad (3)$$

where k is the number of parameters, $V(y)$ is the output variance, $V_i$ is the variance explained by only parameter $x_i$, and $V_{ij}$ is the variance explained by the interaction between parameter $x_i$ and parameter $x_j$. For dependent inputs, this decomposition can be modified based on the joint distribution of the inputs (Chastaing et al. 2012).

The calculation of first-order and second-order variances are shown as examples here:

$$V_i = V_{x_i}[E_{x_{\sim i}}(y|x_i)], \quad (4)$$

$$V_{ij} = V_{x_i,x_j}[E_{x_{\sim i,j}}(y|x_i,x_j)] - V_i - V_j, \quad (5)$$

where $E_{x_{\sim i}}(y|x_i)$ is the expectation of output *y* when controlling factor $x_i$ at a certain value, $V_{x_i}[E_{x_{\sim i}}(y|x_i)]$ represents the variance of all the expectations when controlling $x_i$ at different possible values. $V_i$ shows the influence of factor $x_i$ only. For example, a $V_i$ close to zero means changing $x_i$ does not change the output greatly so that $x_i$ is not important. Similarly, $E_{x_{\sim i,j}}(y|x_i, x_j)$ is the expectation of output *y* when controlling $x_i$ and $x_j$, and $V_{x_i,x_j}[E_{x_{\sim i,j}}(y|x_i, x_j)]$ represents the variance of all the expectations when controlling $x_i$ and $x_j$. After subtracting the individual effect from $x_i$ and $x_j$ ($V_i$ and $V_j$), we get $V_{ij}$ that represents the interaction effect of parameter $x_i$ and parameter $x_j$. The same rules apply to higher-order terms.

Normalizing these variances by the output variance $V(y)$ leads to first-order and second-order Sobol' sensitivity indices.

The total-order Sobol' sensitivity index represents the overall effect of a certain parameter (Homma and Saltelli 1996). For example, the total-order effect of $x_i$ ($T_i$) is the sum of all the terms in equation 3 with *i*. This sensitivity index is more commonly used to quantify the overall importance of a certain parameter to the output variance. Its calculation is shown in equation 6:

$$T_i = E_{x_{\sim i}}[V_{x_i}(y|x_{\sim i})]/V(y), \quad (6)$$

where $V_{x_i}(y|x_{\sim i})$ is the variance when controlling all other parameters at there certain values but $x_i$. The total-order sensitivity index is the expectation of these variances normalized by the output variance.

Sobol' analysis often requires a large sample size to numerically converge (Saltelli 2002; Herman et al. 2013). This becomes the main obstacle of Sobol' method for high-dimensional or slow models. We choose the "sensobol" package in R as the tool for Sobol' analysis. Unlike most packages, this implementation enables the user to separate the model evaluation and sensitivity analysis from the entire analysis (Puy et al. 2021). This simplifies the recording of the computational costs in each step. We adopt the Sobol' quasi-random number sampling design as it often has the highest efficiency (Puy et al. 2021).

We adopt the following convergence criterion for sensitivity indices: a Sobol' sensitivity index converges if the range of its 95% confidence interval is less than 0.05 (Sarrazin et al. 2016). For each test model, we go through a series of increasing sample sizes until all the total-order sensitivity indices converge. The total computational time of the standard Sobol' method is the sum of model evaluation time and sensitivity analysis time.

### 3.2. Sobol' with Kriging emulator

Kriging (also known as Gaussian process regression) is a popular interpolation method that uses a weighted average of sample data to estimate the unknown data (Krige 1951; Oliver and Webster 1990). Kriging is an emulator that can evaluate the estimated output of any given input. The weights are estimated from the sample outputs and the estimated correlation structure among the samples.

The best linear unbiased predictor of Kriging interpolation includes a deterministic part (or mean function) and a stochastic part (Echard et al. 2011). The stochastic part is a Gaussian process, which means any finite collection of elements has a joint normal distribution. This feature means the estimated variance can reflect the data uncertainty. In this study, we adopt simple Kriging, which assumes the mean function is a constant in the entire domain.

We use the "GPfit" package in R to perform Kriging emulation and adopt the default settings (MacDonald et al. 2015): all the input parameters lie between zero and one; the correlation function is an exponential correlation with a power of 1.95. When samples are too close to each other, it is possible for the Gaussian process emulator to become numerically unstable (Deutsch 1996). The "GPfit" package avoids this potential issue by implementing a more computationally stable approach (Ranjan et al. 2011; MacDonald et al. 2015).

For emulation-based methods, we need to know when to stop sampling when building the emulator. One general recommendation for Gaussian process fitting is to start with a sample size of ten times of model dimension (Loeppky et al. 2009). However, this may not be enough for high-dimensional models. We adopt a stopping criterion that can be easily quantified for any given emulation-based method. First, we select a large set of training inputs by Latin Hypercube Sampling (LHS) (McKay et al. 1979). We pick the training input size to be 20,000 following a previous study (Echard et al. 2011). Then we fit an emulator with a different, much smaller set of LHS samples (thus the choice of the training input size does not impact the computational cost of emulator fitting). The Kriging emulator gives the best prediction and the mean-squared error associated with it for each input. We check the root-mean-squared errors for the 20,000 training points given by the fitted emulator. If all the root-mean-squared errors are less than one, we treat the emulator as accurate enough. Note that this is a subjective choice for a stopping criterion. We may also choose other criteria such as comparing the best estimates with the true responses or using other numerical thresholds. Different stopping criteria could influence the required sample size greatly, thus influencing the required computational time. We discuss this topic in section 5.

After sampling for the Kriging emulator, we record the model evaluation time and emulator fitting time. Typically, high-dimensional models require more time in the emulator fitting step. Finally, we adopt the emulator as our new test model and apply Sobol' analysis with the same procedures described in section 3.2 on it. This enables us to estimate the total computational time for the sensitivity analysis part calculated as the sum of the adjusted model evaluation time, emulation time, and sensitivity analysis time.

### 3.3. Sobol' with adaptive Kriging

Using emulation-based methods can reduce the required sample size over the original model compared with standard Sobol' sensitivity analysis. However, the required sample size of emulation can still be relatively large for models with high dimensions or irregular responses.

Adaptive sampling is an approach widely used in many modeling applications to decrease the sample size (e.g. Echard et al. 2011; Garud et al. 2017; Lookman et al. 2019; Wei et al. 2021). Adaptive sampling often involves a certain active learning strategy and chooses the next optimal sample based on it. For example, in model reliability analysis, adaptive sampling means finding the next location where the model is the most unstable or uncertain. Similarly, for the Kriging emulation, we can use an adaptive Kriging approach combined with Monte Carlo sampling (AKMCS) to minimize the number of model evaluations (Echard et al. 2011).

Generally, AKMCS begins with preparing a large set of training input samples. Similar to the first step of the Kriging emulator described above, we take 20,000 LHS samples. Then we randomly select a small subset of them and evaluate the model to get their outputs. We use these samples to fit a Kriging emulator as described in the previous section. For the rest of the unselected training samples, we use the fitted Kriging emulator to get their best estimates and the root-mean-squared errors. Based on these results, we use a learning function to determine the next optimal sample.

Depending on the goal of the analysis, we can define different learning functions. An example goal can be identifying which input parameters lead to extreme outputs close to a certain threshold with the learning function (Echard et al. 2011):

$$U(x) = \frac{G(x) - l}{\sigma}, \tag{7}$$

where $U(x)$ is the learning function for a given input $x$, $G(x)$ is the best estimate of the Kriging prediction at the input $x$, $l$ is the numerical threshold of the extreme output, $\sigma$ is the Kriging root-mean-squared error at the input $x$. In each iteration, the next optimal sample is the one with the least $U(x)$ among the remaining samples. The Kriging fitting stops when the minimum $U(x)$ is less than two. The main idea behind this learning function is choosing the sample with a large uncertainty or close to the threshold.

We modify the above learning function and adopt the root-mean-squared error as a simplified learning function because we do not focus on a specific numerical threshold in our example:

$$U(x) = \sigma, \tag{8}$$

where again $U(x)$ is the learning function for a given input $x$, $\sigma$ is the Kriging root-mean-squared error at the input $x$. When choosing the next sample, we pick the sample with the largest root-mean-squared error, which means the emulator has the largest uncertainty near this location. We stop the Kriging fitting if the largest root-mean-squared error is less than one. We choose this learning function and stopping criterion to make it consistent with the non-adaptive Kriging emulation.

### 3.4. Sobol' with Bayesian Adaptive Spline Surface (BASS)

BASS is an emulation-based method in which the emulator is fitted through Reversible Jump Markov Chain Monte Carlo (RJMCMC) (Green 1995; Francom et al. 2018; Francom and Sansó 2020). This method works can handle functional response models, but we only focus on its basic application on models with single numerical output. The BASS emulator is based on the spline regression. The emulator is a sum of multiple basis functions, where each basis function is a product of piecewise polynomials. Building the emulator requires the estimation of a series of hyperparameters related to these basis functions.

The RJMCMC stochastic method samples the posterior distributions of these hyperparameters adaptively [(Francom et al. 2018)](). RJMCMC can be used to parallelize the computation using multiple chains with relative ease [(Green 1995)](). We use only the default single-chain computation here because we do not consider parallelization in this study. The posterior samples of these hyperparameters form a posterior distribution of the BASS emulator. This can be treated as the uncertainty of the emulator. We then perform the sensitivity analysis for each emulator sample and treat the ensemble as the uncertainty of sensitivity analysis indices.

BASS employs an analytical approach to directly calculate any order Sobol' sensitivity indices based on the emulator [(Francom et al. 2018)](). This means we do not need to take samples and evaluate the emulator for BASS method. This saves the computational cost of evaluating the emulator.

The BASS emulator structure and RJMCMC process require pre-specified settings. In this study, we adopt the default basis function structure setting of the "BASS" package in R. Specifically, we adopt the maximum degree of interaction as three and the maximum amount of basis functions as 1,000. The RJMCMC process does not have a default setting. We set the chain length to 500,000, with the first 100,000 samples being the burn-in period. We keep the results for every 1,000 samples after the burn-in period.

We adopt the same sampling strategy and stopping criterion as used above for the Kriging emulator method (section 3.2). We sample 20,000 LHS samples as training data [(McKay et al. 1979)](). We begin fitting the emulator with LHS samples with a size of ten times the dimension of the parameter space. After each BASS emulation, we stop sampling if all the root-mean-squared errors of the 20,000 training points are less than one. Otherwise, we increase the sample size by the number of model dimensions. Note that this approach only increases the sample size and keeps other settings the same. The word "adaptive" in the BASS method's name refers to adaptive parameter proposal in the RJMCMC step instead of adaptive sampling (and not to adaptive sampling as discussed, for example in the summary of the AKMCS method in section 3.3). We do not optimize the total computational time by adjusting the sample size and the RJMCMC chain length.

## 4. Results

We analyze the results from two aspects: sensitivity indices and total computational time.

## 4.1. Sensitivity analysis results

We use the results of the total-order sensitivity for the third parameter in the five-dimensional test model as an example to quantify the performances of these methods. We choose the five-dimensional model because it is relatively fast so calculating the sensitivity indices at different sample sizes is computationally feasible. We choose the third parameter's total-order sensitivity as an example to visualize because it is the dominant effect in this model. This allows us to analyze the convergence process more clearly. We use five different random seeds for the emulation-based methods. Because the standard Sobol' method samples with the quasi-random Sobol' sequence method, its results do not change with random seeds.

Theoretically, every third parameter will have the same and the largest sensitivity index in our test models. For the five-dimensional model, the third parameter's total-order sensitivity index should be close to one.

All four methods lead to converged total-order sensitivity indices close to one but with different sample sizes (Figure 2). The arrows in each panel of figure 2 represent the endogenously determined converged sample size. The line at the bottom labels the required sample size for each method. The standard Sobol' method requires a sample size much larger than the other methods examined here. The sensitivity index is not stable even after 10,000 samples. The approximation errors between each emulation-based method and the standard Sobol' method all decrease to small values quickly with less than 10 samples (S1 figure). More specifically, the AKMCS method converges with about 40 samples, the Kriging method converges with about 80 samples, and the BASS method converges with about 50 samples. For the emulation-based methods, setting different random seeds leads to similar results, which means these methods are robust.

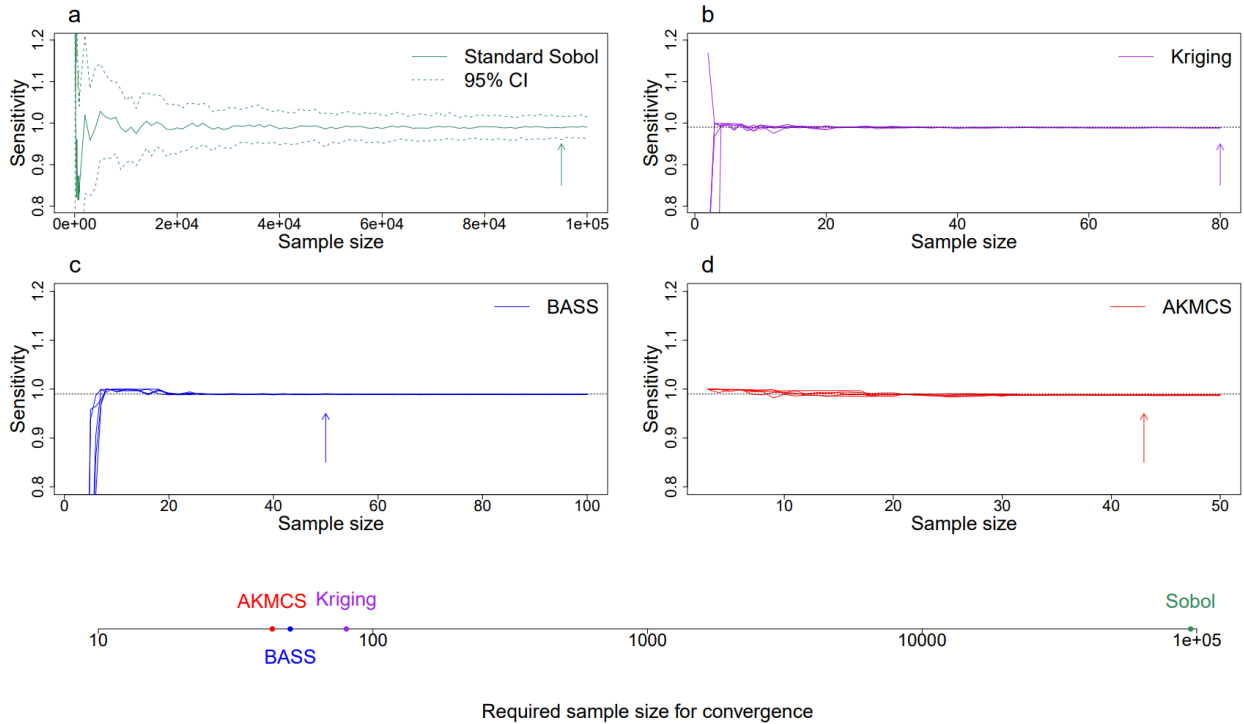

**Figure 2: Trace plot of the total-order sensitivity under each method and each random seed.** Shown are the sensitivity of the third parameter in the 5-dimensional model. Each panel represents a method. Each line within the same panel represents a random seed. The horizontal dashed line represents the best estimate given by the Sobol' method. The arrow in each panel represents the sample size where we stop sampling. For the standard Sobol' method (Figure 2 panel a), the results are independent of the random seeds, and the dashed lines represent the 95% confidence interval of the total-order sensitivity index in panel a.

### 4.2. Computational time results

We use the total computational time without any parallelization as our measure of the required computational costs. Different model characteristics have different impacts on the total computational time. The sample size and the model speed mainly influence the model evaluation time. Both effects are linear. The model dimensionality mainly influences the emulation time, and this effect is strongly nonlinear. The time used in sensitivity analysis depends on a variety of factors such as the sample size, model dimension, and model output behavior.

These relationships can lead to a complex relationship between each method's performance and the model characteristics. Emulation-based methods save model evaluation time due to smaller sample sizes of the original model, but need extra emulation time, especially for high-dimension models. The AKMCS method has the smallest required sample size due to adaptive sampling but needs to fit the Kriging emulator repeatedly. Hence the emulation-based methods outperform the standard Sobol' method for slower models, especially low-dimensional slow models (S2 - S4 figures).

We compare the total computational wall time of each method under each model scenario. The choice of the fastest method differs under different model scenarios (Figure 3). Standard Sobol' uses the least computational time for computationally cheap models (less than 0.1 seconds). The BASS method is the fastest for high-dimensional and computationally expensive models. The kriging method is the fastest for low-dimensional models with a medium run time. The AKMCS method is the fastest for low-dimensional models with an extremely long run time.

The overall pattern is consistent with our initial hypothesis that emulation and adaptive sampling methods are faster than the standard Sobol' method when the model running time is long due to the reduction of original model evaluation time. Emulation can be unnecessary if the original model is fast enough. For slow models, using the fastest emulation-based method can save the computation costs by two to three orders of magnitude (S5 figure). For models running within 1ms, the standard Sobol' method is the fastest with the only exception being the two-dimensional 1ms model. Kriging emulation outperforms standard Sobol' slightly because the emulator is computationally cheaper under low dimensions. For higher-dimensional models, the Kriging emulation time is too long. For the slower models, the BASS method is the fastest for most high-dimensional models due to its relatively fast emulation and numerical sensitivity analysis. The BASS method also performs well for most other model scenarios as its computational cost is often close to the fastest model (S6 figure). For low-dimensional models, Kriging emulation is faster than BASS emulation. Moreover, for low-dimensional slow models, adaptive sampling in AKMCS saves sampling time. This explains the region of Kriging and AKMCS methods.

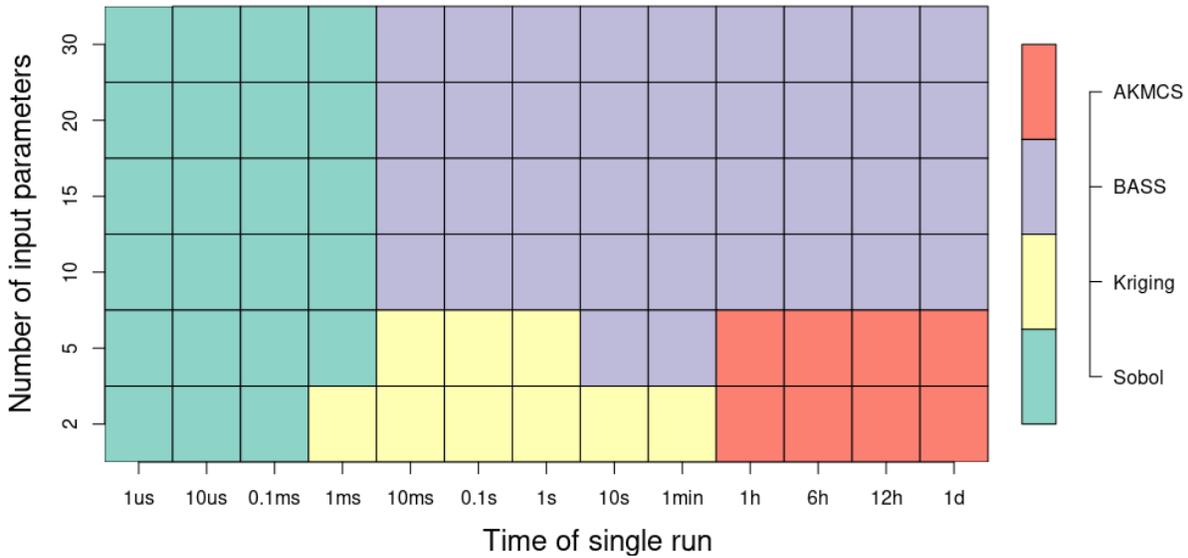

**Figure 3: The fastest method choice for each test model scenario.** The x-axis labels the wall time of different models. The y-axis labels the dimension of the parameter space of different models. The colors inside the grids represent the choice of the fastest method (the one with the shortest total wall time) for the corresponding model scenario (AKMCS = Adaptive Kriging with Monte Carlo Sampling, BASS = Bayesian Adaptive Spline Surface). As model evaluation time increases, the fastest method shifts to emulation-based methods. The BASS method is the fastest for most considered high-dimensional and slow models.

Slow or high-dimensional models still need considerable computational resources for a sensitivity analysis even with the fastest method (Fig 4). Figure 4 shows the approximation of the total computational time for each test scenario. The number in each grid is the order of magnitude of the computational time normalized by the lower-left grid (round down to the closest integer). The letters represent the corresponding fastest methods (S = standard Sobol'; K = Kriging; B = BASS; A = AKMCS), which are the same as in figure 3. The absolute total computational time of the lower-left grid is about 0.1s. This means the time required for the 30d-1day model takes between $10^8 \times 0.1s \approx 3.8$ months and at most $10^9 \times 0.1s \approx 38$ months. This is often infeasible without high-performance or parallel computation.

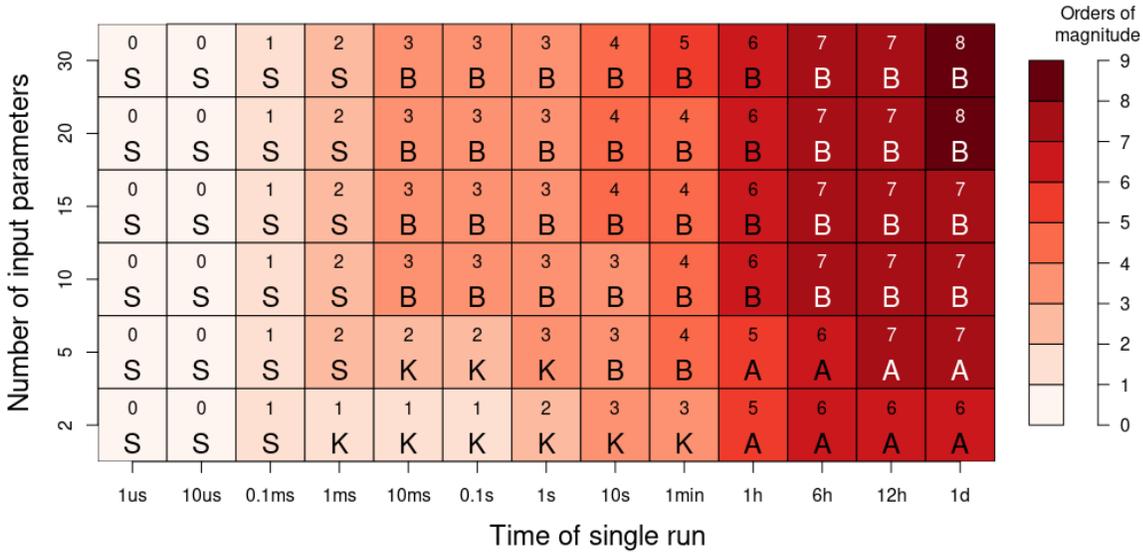

**Figure 4: The total computational time for each test model scenario.** The numbers inside the grids represent the orders of magnitude (round down to the closest integer) of the computational time for that grid normalized by the lower-left grid's computational time.

## 5. Caveats and Open Research Questions

We perform a set of relatively simple numerical experiments to help to inform the decision on which global sensitivity analysis method to use depending on the dimension of the model parameter space and the model evaluation time. Choosing a relatively simple approach helps with the objective of transparency, but also has some disadvantages. Here we briefly discuss some important choices and explain the reason for our choices focusing on (i) the test model structure, (ii) the sensitivity analysis method, (iii) the emulation stopping criterion, and (iv) parallelization in computation.

The first choice is the test model structure. We choose a polynomial deterministic model with a single output for simplicity. For models with stochastic processes, multiple outputs, functional outputs, or non-numerical outputs, most of the sensitivity analysis methods we use in this study need to be modified [(Gunawan et al. 2005; Pianosi and Wagener 2015; Lam 2016; Francom et al. 2018)](). The modification of these methods is a complex topic that is beyond the scope of this paper. Polynomial models with relatively low orders often behave well, which means their response surfaces have no sudden changes or discrete outliers.

The coefficients (repeated 1, 10, and 100 patterns) before each term help us obtain a rough estimate of the theoretical sensitivity analysis indices. For polynomial models, the coefficients largely determine the sensitivity of each parameter. Thus, we design a model structure with very different coefficients before each term. This helps confirm whether all the methods lead to sensitivity analysis results as expected.

We do not expect the same numerical results for different model applications even with the same dimension and evaluation time. The required sample size of the sensitivity analysis depends on model behaviors. The total computational time depends on the specific computational environment. Hence, we expect the required sample size and the total computational time to increase for more complex models. However, the relative trend of the total required time should be similar, which means we expect to see a map of the fastest method choice similar to figure 3. This is because the relative relationship between each method's performance and model dimension and model evaluation time does not change.

Avenues for future research include testing different types of models (i.e. models with dynamic or stochastic outputs) (Gunawan et al. 2005; Lamboni et al. 2011; Pianosi and Wagener 2015). Risk management involving complex dynamics sometimes requires these models (Cano et al. 2016; Vogel 2017).

The second choice is the sensitivity analysis methods we consider. We pick the Sobol' method and Sobol' sensitivity indices as the basis of the comparison because: (1) it is the most commonly used and well-studied method for global sensitivity analysis; (2) using one method as the basis enables us to compare the results among different methods conveniently. More sensitivity analysis methods beyond global variance-based sensitivity analysis could be applied for different research goals.

The third choice is the stopping criterion used for the emulation-based methods. This influences the required sample size greatly. A less strict stopping criterion can save time in sampling but lead to less precise sensitivity analysis results. Users need to balance the tradeoff between the available computational resources and the expected accuracy of sensitivity analysis. For example, different stopping criteria with faster convergence rates can be applied if the research goal of the sensitivity analysis is parameter ranking or screening (Sarrazin et al. 2016).

The fourth choice is that we do not consider computational parallelization. In reality, evaluating model outputs of independent samples can be easily parallelized by multiple cores or machines. This decreases the computational time in the model evaluation part for the Sobol', Kriging, and BASS methods. The impact of parallelization is linear and can be treated as the decrease of single model evaluation time. Parallelization is much more complex for the AKMCS method because every new sample is dependent on previous samples. The BASS method further supports parallelization in its RJMCMC algorithm. Thus, we expect the BASS method to benefit the most if parallelization is available, while the AKMCS method faces considerably more challenges for parallelization.

Considering these choices and the fact that different machines have different computational power, our results can only provide a rather approximate estimate of computational costs for a specific analysis that differs from this study. Our study is designed to provide: (i) basic guidance on how to choose the fastest global sensitivity analysis method with a given model dimension and model evaluation time and (ii) a rough sketch of how the required computational time changes among different methods.

## 6. Conclusion

We analyze the performance of different methods and sampling choices for global sensitivity analysis. The standard Sobol' method using Sobol' sequence sampling is the basis of all methods. The three other methods refine this approach in different ways. The Kriging method and the BASS method build an emulator by Latin Hypercube sampling. The AKMCS method builds a Kriging emulator by an active learning sampling method. We adopt the same stopping criterion for all emulation-based methods to control the experiments.

We consider a test model with a polynomial structure that can be easily expanded into different dimensions. The coefficients before each term in the test model help us have a rough estimation of the true sensitivity indices. Different combinations of model dimensions and evaluation times form our test model scenarios.

All four methods can give reasonable sensitivity indices, but with drastically different sample-size requirements. Standard Sobol' method requires a much larger sample size than emulation-based methods. Though emulation-based methods use a smaller sample size, they require additional computational resources in building the emulator, especially for high-dimensional models. The AKMCS method needs to repeat the emulation process in each learning step.

We find that the standard Sobol' method is the fastest for computationally cheap models, and the BASS method is the fastest for high-dimensional slower models. For models with medium evaluation time and low dimensions, the Kriging method is the fastest. For models with extremely long evaluation times and low dimensions, the AKMCS method is the fastest.

We provide general guidance on how to choose the fastest global sensitivity analysis method depending on model dimensions and evaluation times. Real applications often deal with more complex models, but we hypothesize that the basic relationship between the computational cost and model characteristics to be similar to our conclusions. Finding a global sensitivity analysis method for slow models with a high dimensional input parameter space still poses considerable challenges.

## Author contributions:

Haochen Ye performed the numerical analyses and wrote the first draft.

All authors contributed to the study design, organization, and editing of the paper.

## Data and code availability:

The entire analysis is performed in R. The analysis code is available through GitHub:

https://github.com/yhaochen/Sensitivity

## Acknowledgments:

This work was co-supported by the U.S. Department of Energy, Office of Science, Biological and Environmental Research Program, Earth and Environmental Systems Modeling, MultiSector Dynamics under Cooperative Agreements DE-SC0016162 and DE-SC0022141, the Penn State Center for Climate Risk Management, and the Thayer School of Engineering at Dartmouth College. We thank the support of the Program of Coupled Human and Earth Systems (PCHES), the PCHES research team, and the Keller research group for discussions. Special thanks to Murali Haran and Jim Kasting for their suggestions and advice, and to Matthew Lisk and Atieh Alipour for code checking and replication. Any opinions, findings, conclusions, and recommendations expressed in this material are those of the authors and do not necessarily reflect the views of the funding entities.

**Supplementary Materials:**

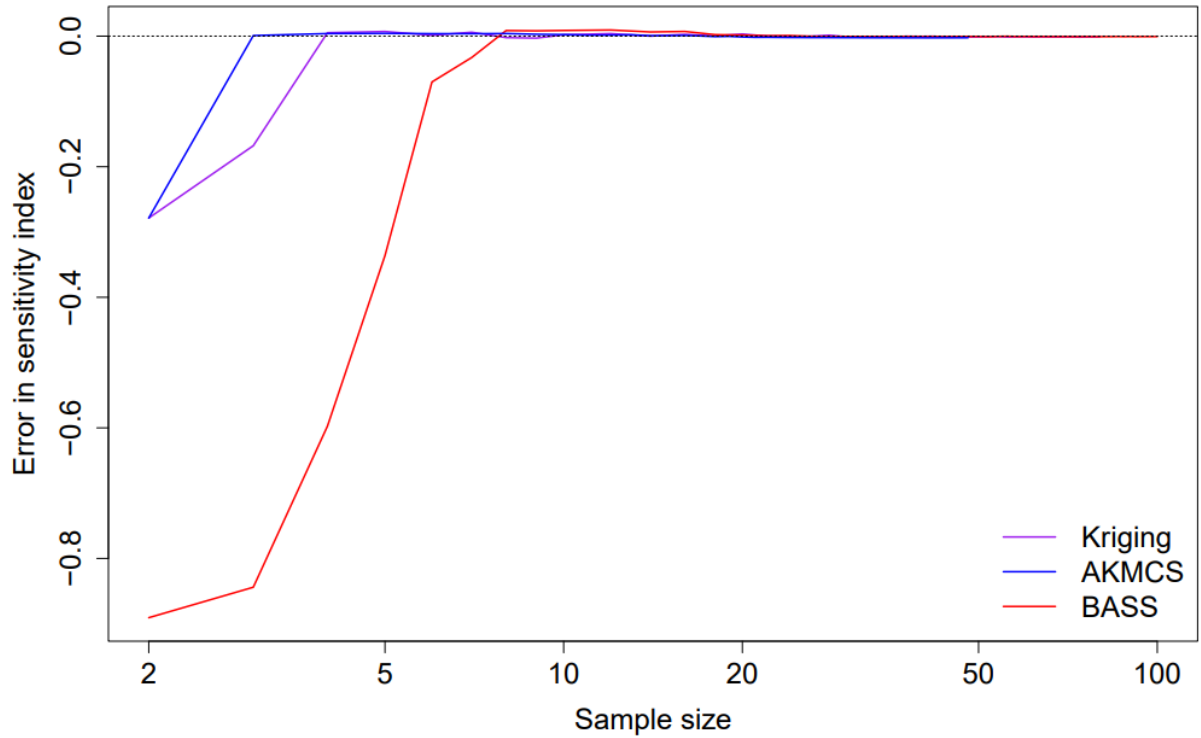

**Supplementary Figure 1: The approximation error in the total order sensitivity index for each method.** The same as Figure 2, this figure uses the third parameter in the 5-dimensional test model as an example. The y-axis is the difference between the total-order sensitivity index from the standard Sobol' method (the best estimate when converged) and the total-order sensitivity index from emulation-based methods (the average of the best estimates in five random seeds). We see the errors decrease quickly to around zero within 10 samples for all three emulation-based methods, where the AKMCS method converges the fastest and the BASS method converges the slowest.

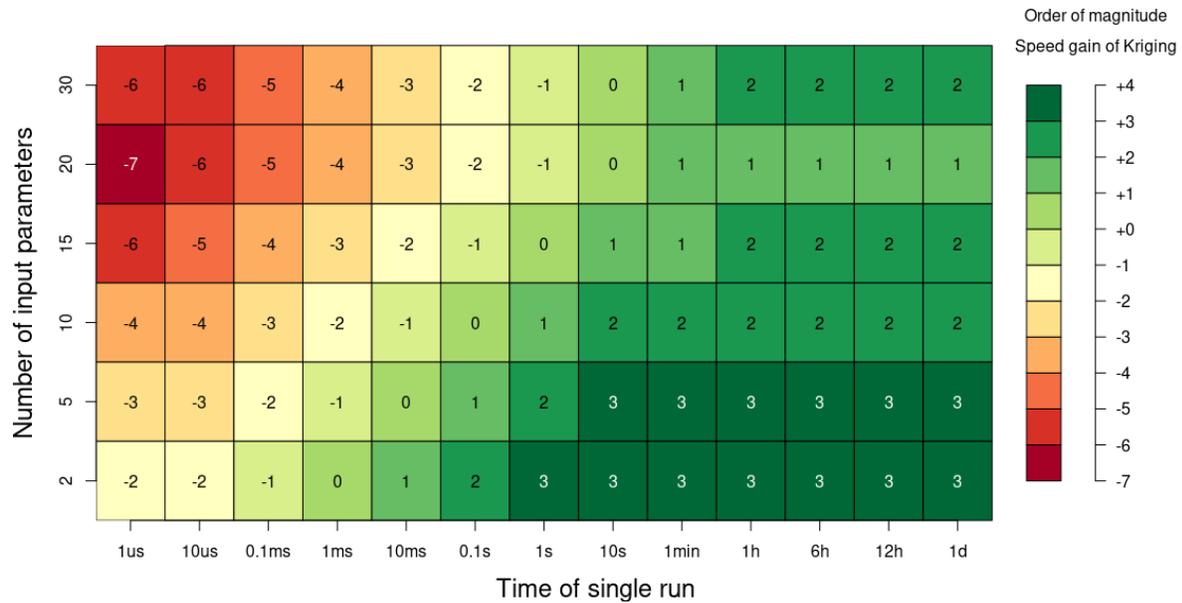

**Supplementary Figure 2: Speed gain of the Kriging method compared with the standard Sobol' method.** Within each grid, the number represents the orders of magnitude of the computational time ratio between the standard Sobol' method and the Kriging method (round down to the closest integer). For example, a number -2 means the Kriging method is between 10 times and 100 times slower than the standard Sobol' method. The Kriging method outperforms the Sobol' method as model evaluation time gets longer. However, the relative speed gain of the Kriging method decreases as the model dimension increases because of the longer emulation time.

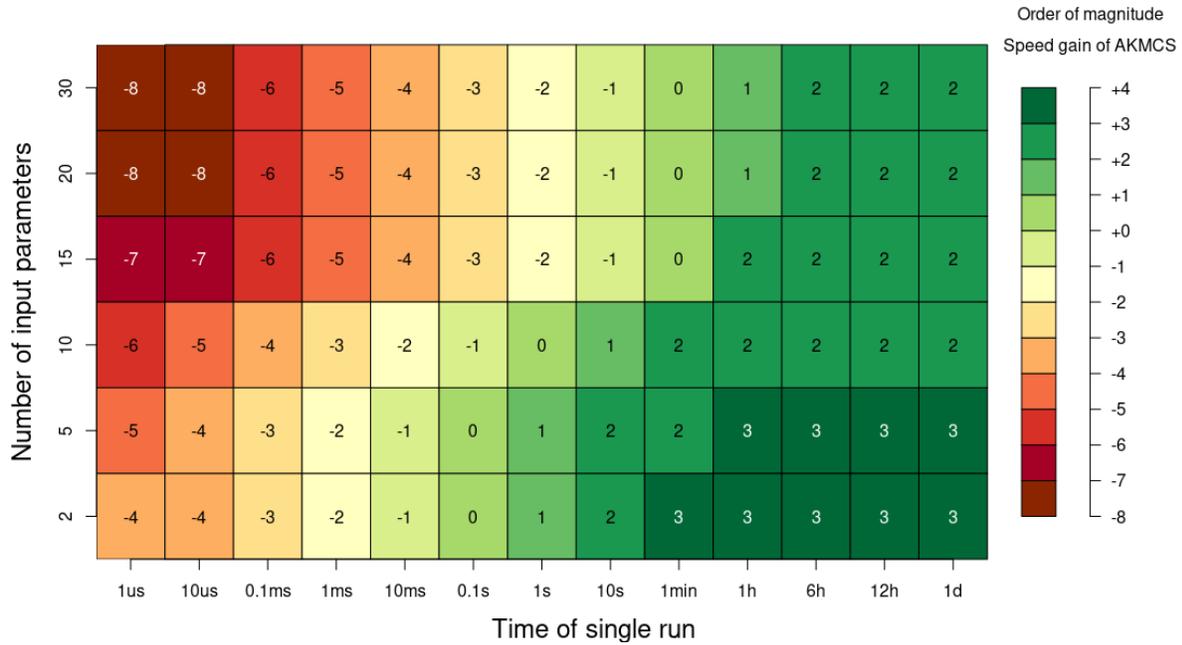

**Supplementary Figure 3: Speed gain of the AKMCS method compared with the standard Sobol' method.** Similar to supplementary figure 2, the numbers represent the orders of magnitude of the time ratio between the two methods (round down to the closest integer). The overall pattern of the AKMCS method's speed gain is similar to that of the Kriging method except that AKMCS's speed gain decreases much more as the model dimension increases.

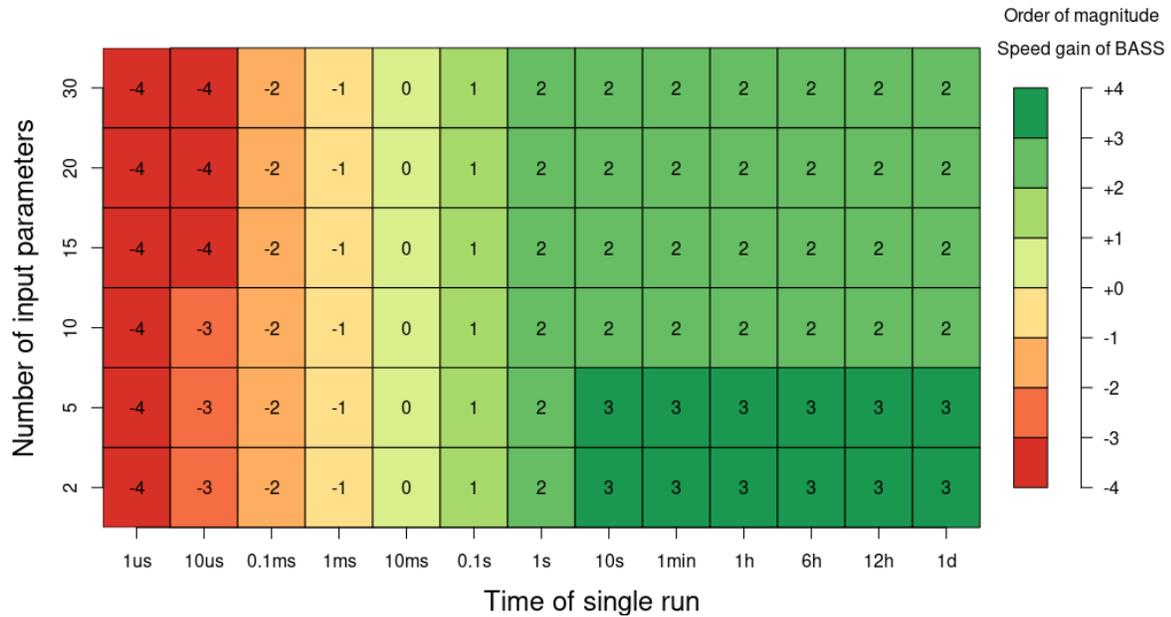

**Supplementary Figure 4: Speed gain of the BASS method compared with the standard Sobol' method.** Similar to supplementary figures 2 and 3, the numbers represent the orders of magnitude of the time ratio between the two methods (round down to the closest integer). The BASS method's speed gain is not influenced by the model dimension greatly.

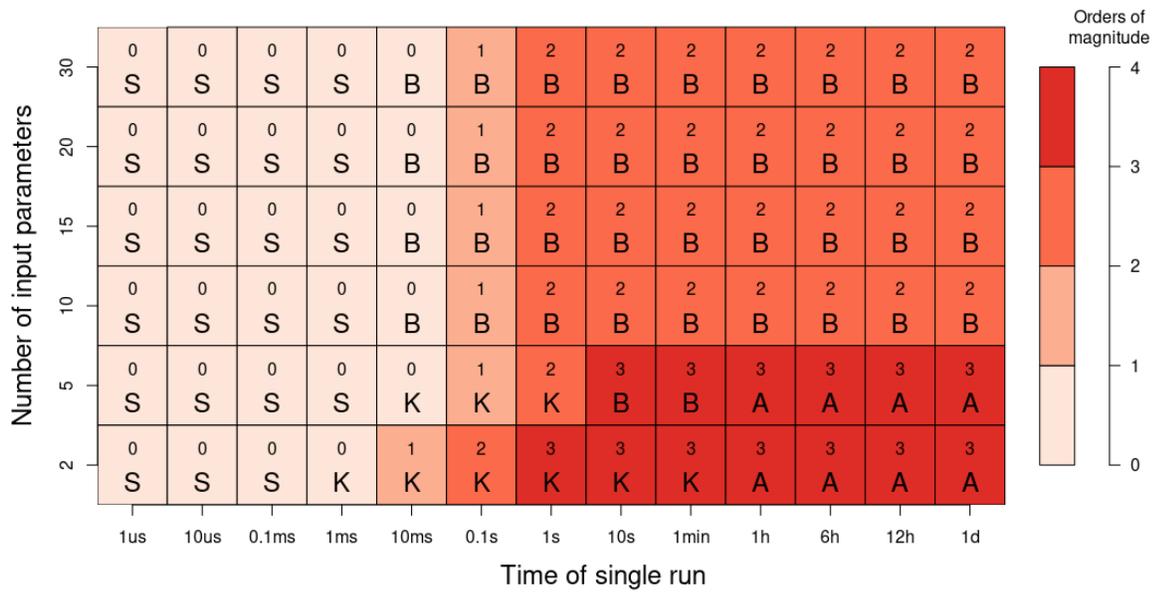

**Supplementary Figure 5: Speed gain by switching from the standard Sobol' method to the fastest method.** In each grid, the numbers are the orders of magnitude of the computational time ratio (round down to the closest integer) and the letters are the fastest method (S = standard Sobol'; K = Kriging; B = BASS; A = AKMCS). For example, for the bottom right corner (the 2-dimensional model that takes one day to run), the fastest method is AKMCS and it is at least 1,000 times faster than the standard Sobol' method.

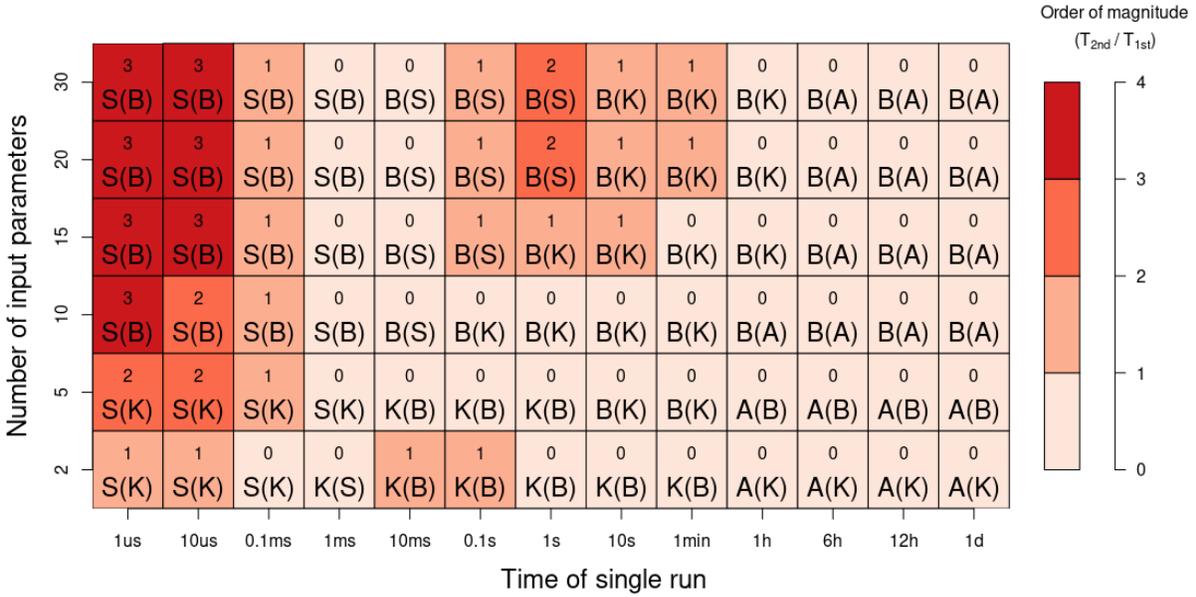

**Supplementary Figure 6: Speed gain by switching from the second fastest method to the fastest method.** We compare the performance difference between the fastest method and the second fastest method for each model scenario. Again, in each grid, the number represents the orders of magnitude of the computational time ratio between the two methods (round down to the closest integer). The letter outside the parentheses indicates the fastest method and the letter inside the parentheses indicates the second fastest method. For example, a number 0 means the two methods' computational times are close to each other (less than 10 times difference). We notice the BASS method is either the fastest or the second fastest for most high-dimensional models, and the AKMCS method is either the fastest or the second fastest for most slow models.